\documentclass[12pt]{article}

\usepackage{epsf}
\usepackage[dvips]{graphicx}

\newcommand{\la}{$\Lambda$}
\newcommand{\al}{$\bar{\Lambda}$}

\begin{document}

\begin{center}

 {\bfseries
LONGITUDINAL POLARIZATION OF $\Lambda$ AND $\bar\Lambda$ IN
DEEP-INELASTIC SCATTERING  AT COMPASS}

\vskip 5mm

M.G.Sapozhnikov\\
on behalf of the COMPASS Collaboration

\vskip 5mm

{\small

 {\it
Joint Institute for Nuclear Research,
Laboratory of Particle Physics,
Dubna
\\
{\it E-mail: sapozh@sunse.jinr.ru }}
 }

\end{center}

\vskip 5mm

\begin{center}
\begin{minipage}{150mm}
\centerline{\bf Abstract} Production of $\Lambda$ and $\bar
\Lambda$ hyperons in deep-inelastic scattering of 160 GeV/{\it c}
polarized muons is under study
 in the COMPASS (CERN NA58) experiment. Preliminary results on longitudinal polarization
of \la~ and \al~ hyperons from the data collected during the 2002
run are presented.
\end{minipage}
\end{center}

\vskip 10mm

The study of longitudinal polarization of $\Lambda (\bar\Lambda)$
hyperons in the deep-inelastic scattering (DIS) can provide
information on the fundamental properties of the nucleon, such as
polarization of strange quarks in the nucleon \cite{Ellis}, and
offers a possibility to determine the  mechanism of spin transfer
from a polarized quark to a polarized baryon
\cite{BJ}-\cite{Ma.00}.

The polarized nucleon intrinsic strangeness model \cite{Ellis}
 predicts negative longitudinal polarization
of $\Lambda$ hyperons produced in the target fragmentation region
\cite{Ell.96,Ell.02}. The main assumption of the model is negative
polarization of the strange quarks and antiquarks in the nucleon.
This assumption was inspired by the results of EMC \cite{EMC.88}
and subsequent experiments \cite{SMC.97}- \cite{E155} on inclusive
deep-inelastic scattering which gave an indication that the
$s\bar{s}$ pairs in the nucleon are negatively polarized with
respect to the nucleon spin:
\begin{equation}
\Delta s \equiv \int\limits_{0}^{1} dx[ s_{\uparrow}(x) -
s_{\downarrow}(x) + \bar s_{\uparrow}(x)-
\bar s_{\downarrow}(x)] = - 0.10\pm0.02.
\end{equation}

Recent analysis \cite{Lea.03} of the available world data in
framework of the next-to-leading order perturbative QCD including
higher twist effects  also concludes that $\Delta
s=-0.045\pm0.007$.

     The polarized strangeness model \cite{Ellis} was successfully
applied to explain the large violation of the OZI rule in the
annihilation of stopped antiprotons and its strong dependence on
the spin of the initial state (for review, see \cite{Sap.03}). The
predictions of the model were confirmed in different processes of
proton-proton, antiproton-proton interactions and lepton DIS.
Specifically, the negative longitudinal polarization of the
$\Lambda$ hyperons at $x_F<0$ predicted in \cite{Ell.96,Ell.02}
was found in the neutrino DIS experiments \cite{NOMAD_l}.

However, the question about polarization of the nucleon strange
quarks has not been solved yet. Recently, after analysis of the
semi-inclusive DIS channels in the LO approximation, the HERMES
collaboration found that $\Delta s =0.028\pm0.033\pm0.009$
\cite{HERMES.04}, i.e. consistent with zero within the errors (for
the discussion of the HERMES result, see
\cite{Kot.03},\cite{Lea.03b}).

The measurement of the longitudinal $\Lambda$ polarization in the
current fragmentation region $x_F>0$ is traditionally related with
investigations of spin transfer from quark to hadron
\cite{BJ}-\cite{Yan.00}. This spin transfer depends on the
contribution of the spin of the struck quark to the \la~ spin.
There are different models of the \la~spin structure.

 In the naive quark model the spin of $\Lambda$ is carried by
the $s$ quark and the spin transfer from the $u$ and $d$ quarks to
$\Lambda$ is equal to zero. It means that the longitudinal
polarization of $\Lambda$ produced in fragmentation of the $u$ and
$d$ quarks is $P_{\Lambda} \sim 0$.

The authors of \cite{BJ}, using $SU(3)_f$ symmetry and
experimental data for the spin-dependent quark distributions in
the proton, predict that the contributions of $u$ and $d$ quarks
to the $\Lambda$ spin are negative and substantial, at the level
of 20\% for each light quark. One might expect that in this model
the fragmentation of the dominant $u$ quark will lead to
$P_{\Lambda}=-0.2$.

In the framework of $SU(6)$ based quark-diquark model
\cite{Yan.00} it is predicted a large and positive polarization of
the $u$ and $d$ quarks in the $\Lambda$ at large Bjorken scaling
variable $x_{Bj}$. Due to this fact the spin transfer to $\Lambda$
should be as large as +1 at $z \sim 0.8-0.9$ (here $z$ is
fractional hadron energy, $z=E_h/\nu$,
 $\nu=E-E'$, $E$ and $E'$ are
lepton energies in the initial and final state).

However the possibility to study real spin transfer from the quark
to baryon at the energies of the current experiments was
questioned in \cite{Ell.02}. It turns out that even at the COMPASS
energy of 160 GeV most of $\Lambda$, even in the $x_F>0$ region,
are produced from the diquark fragmentation. It is predicted that
in the COMPASS kinematics the longitudinal $\Lambda$ polarization
is either $P_{\Lambda}=-0.004$ or $P_{\Lambda}=-0.07$, depending
on the fragmentation model.

Moreover, a large part (up to 30-40\%) of the $\Lambda$, observed
in DIS, comes from decays of heavy hyperons, such as $\Sigma^0$,
$\Sigma(1385)$ and $\Xi$, significantly changing the pattern of
the spin transfer.

More clear situation is with the $\bar{\Lambda}$ production.  The
contribution of \al~ production from diquark fragmentation is
negligible. The background from decays of heavy hyperons is also
absent.  An interesting feature was observed in the E665
experiment at Fermilab \cite{E665}.  It was found that in DIS the
spin transfer to \la~ and \al~  is large and has opposite signs.
Though the statistical errors of the measurement were quite large,
the statistics comprises  only  750 \la~ and 650   \al~  events.

The spin transfer to \la~ and \al~ for the E665 experimental
conditions was calculated in \cite{Ma.00}. It is predicted, that
under standard assumption that unpolarized strange quark
distribution $s(x)$  is the same as $\bar{s}(x)$ for antistrange
quarks, the spin transfer to \la~ should be practically the same
as for \al. Trying to explain the difference in \la~ and \al~ spin
transfer, the authors of \cite{Ma.00} have introduced by hand some
asymmetry in strange-antistrange quark distributions at small
$x_{Bj}$. However, the calculated magnitude of the difference in
\la~ and \al~ spin transfer was still too small to explain the
data \cite{E665}.

We have studied $\Lambda$ and $\bar \Lambda$  production by
polarized $\mu^+$ of 160 GeV/c on a polarized $^{6}$LiD target of
the COMPASS spectrometer constructed in the framework of CERN
experiment NA58. A detailed description of the COMPASS
experimental setup is given elsewhere \cite{COMPASS} and only the
most relevant elements for the present analysis will be given
below.

The beam polarization is $P_b=-0.76\pm0.04$. The polarized
$^{6}$LiD target consists of two oppositively polarized cells, 60
cm long. The target polarization is about 50\%. For this analysis
the data are averaged over the target polarization.

We have used data collected during the 2002 run. The analysis
comprises about $1.6\cdot10^7$ DIS events with $Q^{2} >1~$
(GeV/{\it c})$^{2}$.

\begin{figure}[h]
 \epsfysize=50mm
 \centerline{
\epsfbox{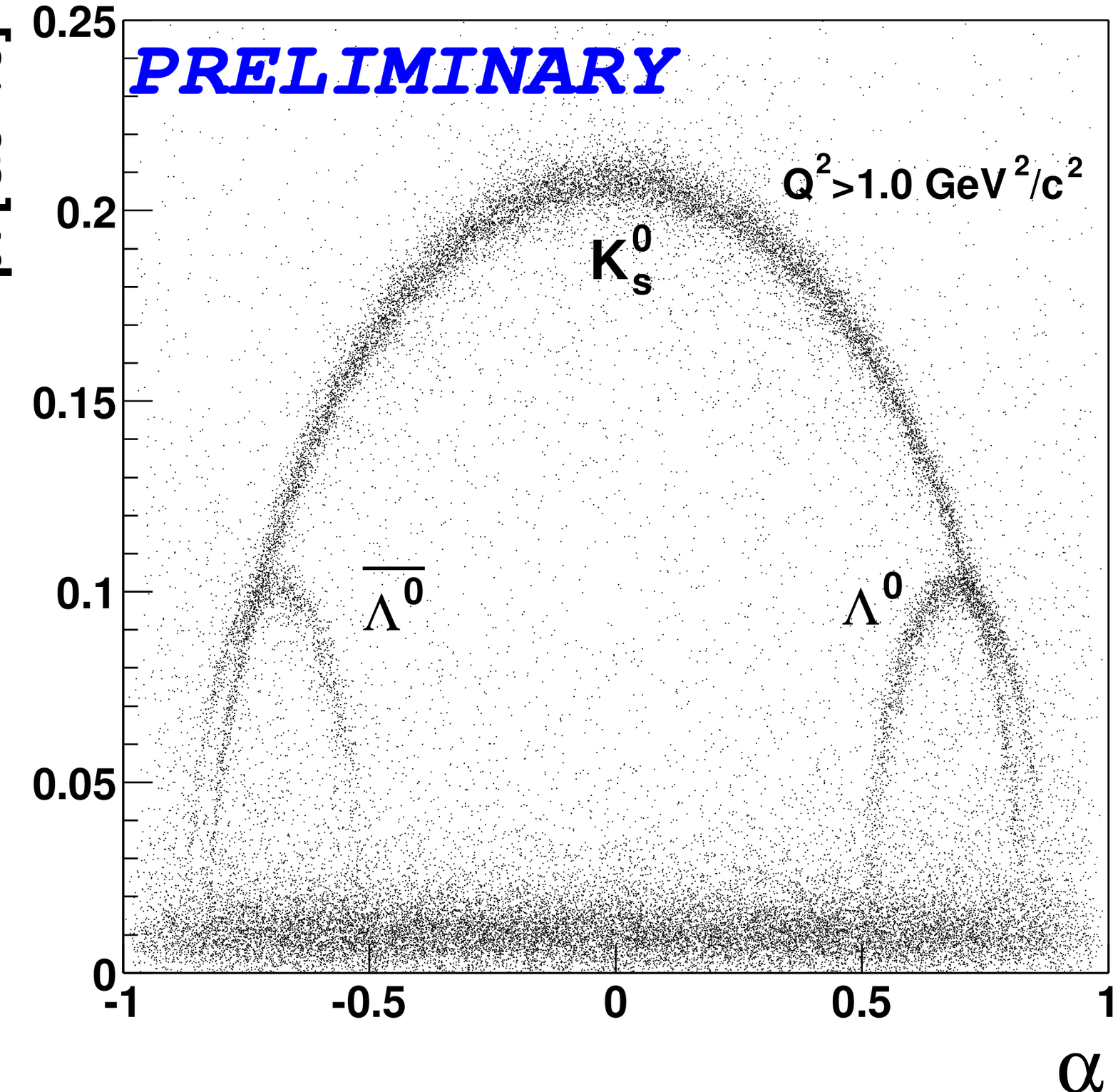}}
 \caption{The Armenteros plot: $p_t$ is the
transverse momentum of the $V^0$ decay products with respect to
the direction of $V^{0}$ momentum,
$\alpha=\frac{p^+_L-p^-_L}{p^+_L+p^-_L}$, where $p_L$ is the
longitudinal momentum of the $V^0$ decay particle. }
\label{arm}
\end{figure}

The $V^{0}$ events ($V^0 \equiv \Lambda$, $\bar\Lambda$  and
$K^{0}_{S}$)  were selected by requiring the incoming and outgoing
muon tracks together with at least two hadron tracks forming the
secondary vertex. The primary vertex should be inside the target.
The secondary vertex must be downstream the both target cells. The
angle between the vector of $V^0$ momentum and the vector between
primary and $V^0$ vertices should be $\theta_{col}<0.01$ rad. Cut
on transverse momentum of the decay products with respect to the
direction of $V^{0}$ particle, $p_t >23$ MeV/c was applied to
reject $e^{+}e^{-}$ pairs from the $\gamma$ conversion seen as the
band at the bottom of the Armenteros plot shown in Fig.\ref{arm}.

The typical elliptical bands from the $K_S^0$, $\Lambda$ and
$\bar\Lambda$ decays are seen in Fig. \ref{arm}. Both $\Lambda$
and $\bar\Lambda$ signals stand out clearly. The large number of
produced $\bar\Lambda$ is a specific feature of the COMPASS
experiment.

The  DIS cuts on $Q^{2} >1~$ (GeV/{\it c})$^{2}$ and $0.2< y <
0.8$ have been used. Here  $y=\nu/E$ is the fraction of the lepton
energy carried out by the virtual photon.

 To analyse $V^0$ events, the so
called bin-by-bin method was used. All angular distributions have
been divided in some bins. In each angular bin the invariant mass
distribution of positive and negative particles is constructed
assuming $\pi^+ \pi^-$, $p\pi^-$ or $\bar{p} \pi^+$ hypothesis.
The peak of the corresponding $V^0$ particle is fitted and the
number of the $V^0$ in this peak is obtained. This number
determines a point in the corresponding bin of the angular
distribution. This procedure allows to construct practically
background-free angular distributions. The total data sample
contains about 9000 $\Lambda$ and 5000 $\bar\Lambda$.

\begin{figure}[h]
 \epsfysize=70mm
 \centerline{
\epsfbox{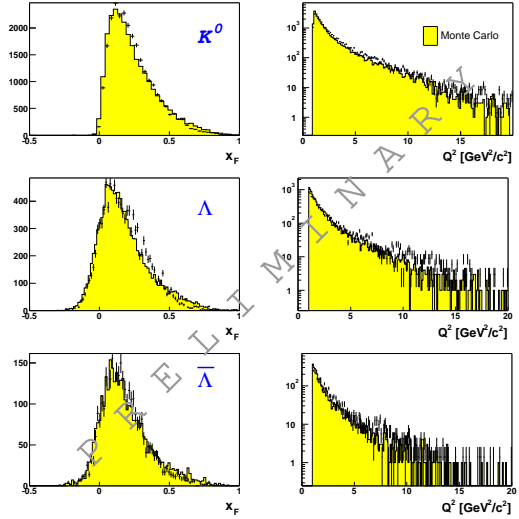}}
 \caption{ The $x_F$ (left column) and $Q^{2}$
(right column ) distributions for
 $K^{0}_{S}$ (upper row), $\Lambda$ (middle row) and $\bar\Lambda$ (lower row).
The experimental data points are shown together with results of
Monte-Carlo simulations (histograms).}
\label{xf}
\end{figure}

The $x_{F}$ and $Q^{2}$ experimental distributions (crosses) of
$\Lambda$, $\bar\Lambda$ and $K^{0}_{S}$ are compared with the
Monte-Carlo simulated ones (hatched histograms) in  Fig. \ref{xf}.
One can see that we are able to access mainly current
fragmentation region. The averaged value of $x_F$ is $<x_F>=0.20$,
whereas for the Bjorken scaling variable it is $<x_{Bj}>=0.02$.
 The mean $\Lambda$ momentum is 17 GeV/c, while decay pion
momentum is 3 GeV/c. The agreement between the experimental data
and the Monte-Carlo simulations is reasonable.

The angular distribution  of the decay particles in the $V^0$~
rest frame is

\begin{equation}
w(\theta)=\frac{dN}{d \cos{\theta}}=\frac{N_{tot}}{2}(1+\alpha P
\cos{\theta}), \label{ideal}
\end{equation}
where $N_{tot}$ is the total number of events,
$\alpha=+(-)0.642\pm0.013$ is \la(\al)~decay parameter, $P$ is the
projection of the polarization vector on the direction of the
virtual photon in the $V^0$ rest frame, $\theta$ is the angle
between the direction of the decay proton for \la~ (antiproton -
for \al, positive $\pi$ - for $K^0$) and the direction of the
virtual photon in the $V^0$ rest frame.

\begin{figure}[h]
 \epsfysize=70mm
 \centerline{
\epsfbox{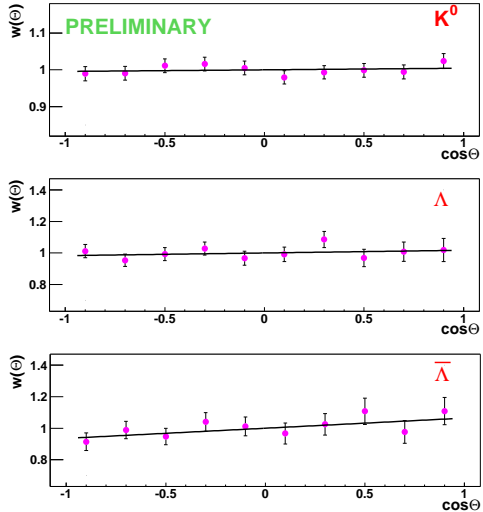}}
\caption{The angular distributions for $K^{0}_{S}$, $\Lambda$ and
$\bar\Lambda$ for all events.} \label{ang}
\end{figure}

Fig. \ref{ang} shows the measured angular distributions
 for all events of the $K^{0}_{s}$, $\Lambda$ and $\bar\Lambda$ decays, corrected for the acceptance.
The acceptance was determined by the Monte Carlo
simulation of unpolarized $\Lambda(\bar\Lambda)$ decays.

One could see that the angular distribution for $K^{0}_{S}$ decays
is flat, as expected. The value of the longitudinal polarization
is $P_K=0.007\pm0.017$. The angular distribution for $\Lambda$
decays for all events, i.e. the averaged other whole $x_F$
interval, is also flat. The value of the longitudinal polarization
is $P_{\Lambda}=0.03\pm0.04(stat)\pm0.04(syst)$. It indicates on
small polarization of $\Lambda$ in DIS processes. The same trend
was observed by the HERMES collaboration \cite{hermes}. The
angular distribution of the $\bar{\Lambda}$ events, averaged over
$x_F$, exhibits some negative polarization
$P_{\bar{\Lambda}}=-0.11\pm0.06(stat)\pm0.04(syst)$.

\begin{figure}[h]
 \centerline{
\epsfysize=75mm
\epsfbox{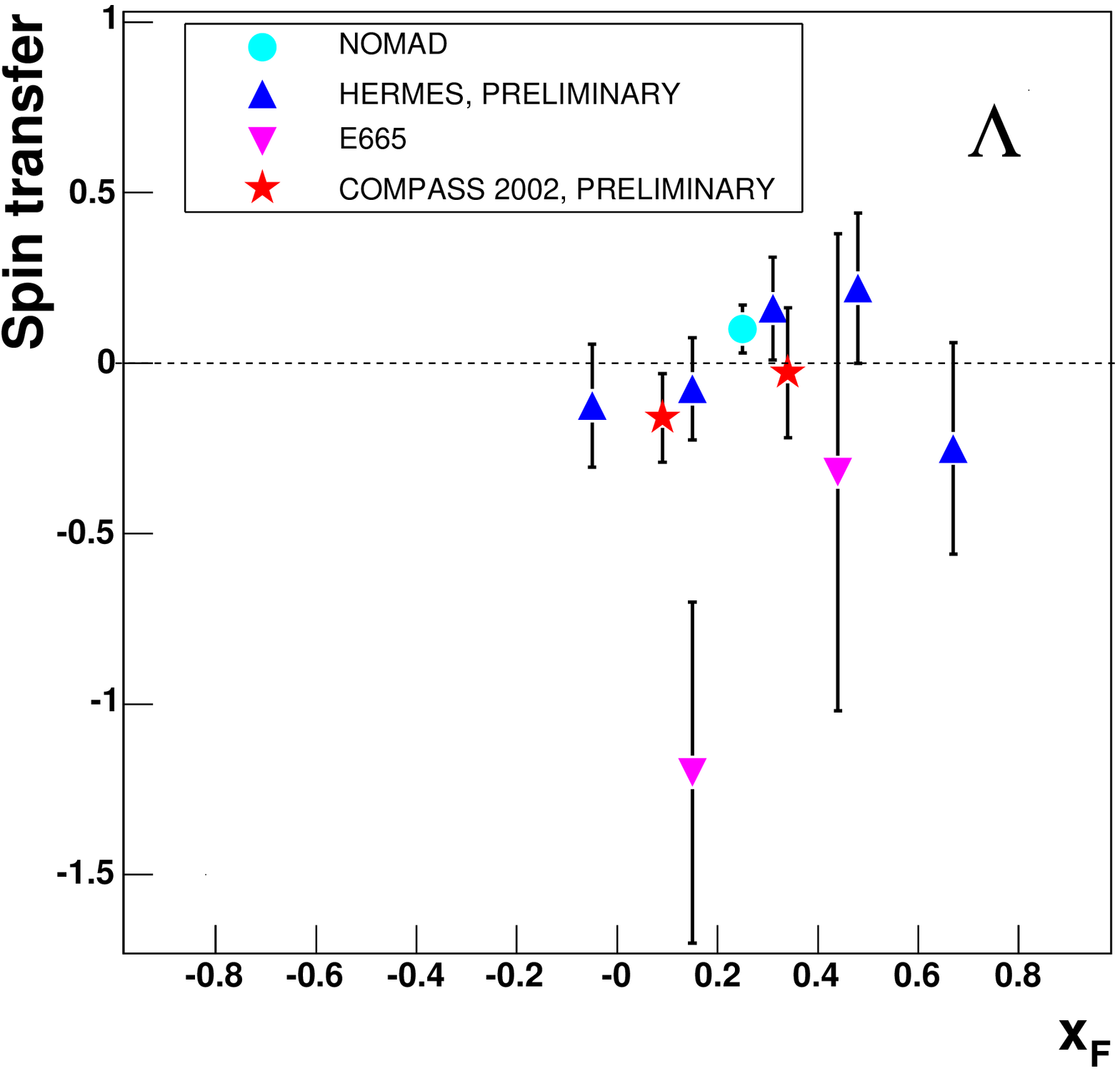}
\epsfysize=75mm
\epsfbox{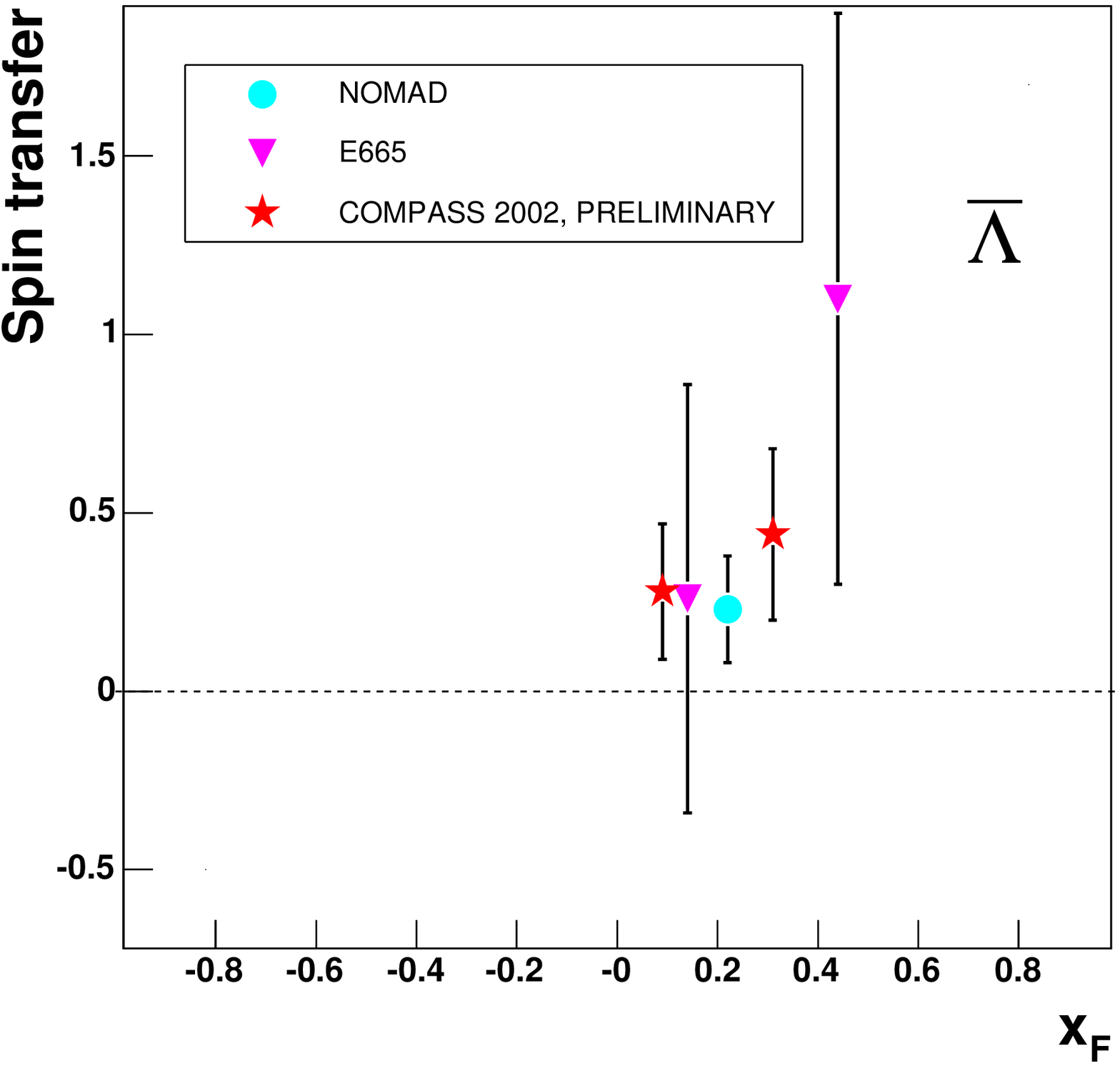}
}
 \caption{Comparison of the
spin transfer to \la~(left) and \al~(right) hyperons measured in
different DIS experiments. } \label{lam2}
\end{figure}

Comparison of the spin transfer to \la~ and \al~hyperons measured
in different DIS experiments is shown in Fig. \ref{lam2}. The spin
transfer $S$ determines which part of the beam polarization $P_b$
is transferred to the hyperon polarization $P$. It is defined as\\
$P=S\cdot P_b\cdot D(y)$, where $D(y)$ is the virtual photon
depolarization factor.

One can see that there is a reasonable agreement between the
COMPASS and world data. There is an indication that the spin
transfer to $\bar\Lambda$ is non-zero and might be different from
$\Lambda$ case. However, for the COMPASS data only statistical
errors are shown. Work to determine the systematic errors is going
on.

The results from our 2002 data demonstrate a good potential of
COMPASS to measure $\Lambda$ and $\bar\Lambda$ polarizations in
DIS. The data samples collected in 2003 and 2004 will
significantly increase the statistics.

\end{document}